\documentstyle[12pt]{article}
\input psfig

\newcommand{\lesssim}{\mathop{}_{\textstyle \sim}^{\textstyle <}}
\newcommand{\text}{\mbox}

\begin{document}


\begin{titlepage}
\begin{center}

\hfill    FERMILAB-CONF-97/418-T\\
\hfill    hep-ph/9712427 \\
\hfill    December 1997

\vskip .5in

{ \Large \bf  Flavor and CP Violations from Sleptons\\ at the 
      Muon Collider} 

\vskip .3in

{\large Hsin-Chia Cheng}

\vskip .3in

{\it Fermi National Accelerator Laboratory\\
     P.O. Box 500\\
     Batavia, IL 60510}

\end{center}

\vskip .3in

\begin{abstract}

Supersymmetric theories generally have new flavor and CP violation 
sources in the squark and slepton mass matrices. They will contribute
to the lepton flavor violation processes, such as $\mu \to e \gamma$, which
can be probed far below the current bound with an intense muon source
at the front end of the muon collider. In addition, if sleptons can
be produced at the muon collider, the flavor violation can occur at
their production and decay, allowing us to probe the flavor mixing structure
directly. Asymmetry between numbers of $\mu^+ e^-$ and $e^+ \mu^-$
events will be a sign for CP violation in supersymmetric flavor mixing.

\end{abstract}

\vskip .5in

\begin{center}
Talk presented at the Workshop on Physics at the First Muon
Collider\\ and at the Front End of a Muon Collider\\
November 6-9, 1997, Fermilab, Batavia, Illinois
\end{center}
\end{titlepage}

\renewcommand{\thepage}{\arabic{page}}
\setcounter{page}{1}
\renewcommand{\thefootnote}{\arabic{footnote}}
\setcounter{footnote}{0}

Weak scale supersymmetry (SUSY) is one of the most attractive
candidates for physics beyond the standard model (SM). The discovery
of superpartners of the SM particles is promising at the planned 
future colliders. If SUSY is discovered, measuring the masses and 
couplings of the superpartners will become the focus of study.
Measurements of the superparticle couplings is essential for
verifying supersymmetry \cite{susytest}, and the superpartner masses
provide information of the origin of SUSY breaking and unification
at high scales \cite{sumrules}. In most SUSY extension of the standard
model, mass matrices of the fermions and their scalar superpartners
are, however, not diagonal in the same basis. New flavor mixing
matrices $W$, analogous to the CKM matrix, will appear at the 
gaugino-fermion-sfermion vertices. These new flavor mixing matrices
may provide clues to the puzzle of the flavor structure, and therefore
should also be important to study. At the muon collider and its front
end, these new flavor mixing effects can be studied both indirectly
through the rare flavor-changing process and directly by the
slepton production if they are accessible. In this talk, we will
discuss the power in probing the SUSY flavor mixings of the muon
collider and compare the indirect and the direct probes. This work
grew out from the studies done with N. Arkani-Hamed, J. L. Feng, and 
L. J. Hall \cite{ACFH1,ACFH2}.

\begin{figure}
\centerline{\psfig{file=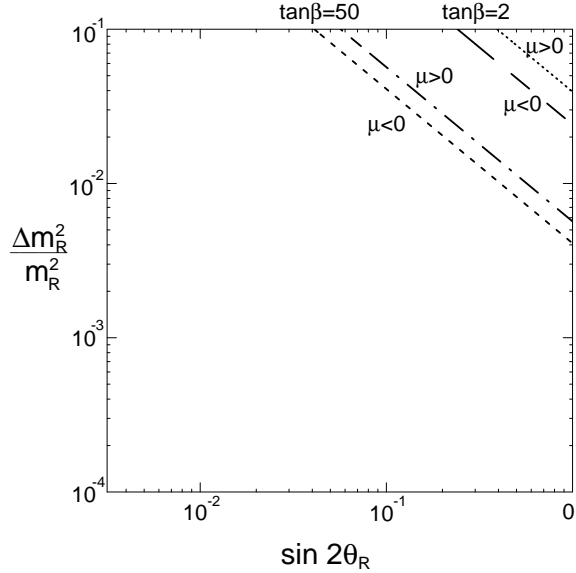,width=0.6\textwidth}}
\caption{Constant contours of current bound on $B(\mu \to e\gamma)
  =4.9\times 10^{-11}$ for the parameters given in the text and
  different $\tan\beta$ and signs of $\mu$.}
\label{muegamma}
\end{figure}

Lepton flavor, although conserved in the SM, is typically violated
in most SUSY extension of the SM, since the scalar partners of 
the fermions must be given mass, and the scalar
mass matrices are generally not diagonal in the same basis as the
fermion masses. When we work in the mass eigenstates of both leptons and
sleptons, the flavor mixing matrices $W$ will appear in gaugino/Higgsino
vertices,
\begin{eqnarray}
\label{convention}
&&\tilde{e}_{Li} {W_L^*}_{i\alpha} \overline{e_{L\alpha}} 
\tilde{\chi}^0
+\tilde{e}^*_{Li} {W_L}_{i\alpha} \overline{{\tilde{\chi}}^0} 
e_{L\alpha}  \nonumber \\
&&+\tilde{e}_{Ri} {W_R^*}_{i\alpha} \overline{e_{R\alpha}} 
\tilde{\chi}^0 
+\tilde{e}^*_{Ri} {W_R}_{i\alpha} \overline{{\tilde{\chi}}^0} 
e_{R\alpha} \, ,
\end{eqnarray}
where the Latin and Greek subscripts are generational indices for
scalars and fermions, respectively. Nontrivial $W$ matrices generate
contributions to the rare flavor-changing processes, such as 
$\mu \to e \gamma$. The $\mu \to e \gamma$ rate is proportional
to (simplifying to 2 generation mixing, $W_{11}=\cos \theta_{12}$,
$W_{12}=\sin \theta_{12}$)
\begin{equation}
 \left( \frac{\Delta
m_{12}^2}{\bar{m}_{12}^2}\sin 2\theta_{12} \right)^2 \, ,
\end{equation}
where $\bar{m}_{ij}^2 = (m_i^2 + m_j^2)/2$ is the average squared
slepton mass, and $\Delta m_{ij}^2 = (m_i^2 - m_j^2)/2 \approx 2m
\Delta m_{ij}$, with $\Delta m_{ij} = m_i-m_j$. It also depends
on other SUSY parameters such as $m_{\tilde{\chi}_0}$, $\bar{m}_{L,R12}$,
$\mu$, $\tan\beta$, and so on. The current bound $B(\mu \to e \gamma) < 4.9
\times 10^{-11}$~\cite{Bolton} puts strong constraints on
$\frac{\Delta m_{12}^2}{\bar{m}_{12}^2}$ and $\sin 2\theta_{12}$.
The constraints on $\frac{\Delta
m_{R12}^2}{\bar{m}_{R12}^2}$ and $\sin 2\theta_{R12}$ (assuming
right mixing only) for $M_1=M_2/2=130 $GeV, $m_{\tilde{l}_R}=200$ GeV,
$m_{\tilde{l}_L}=350$ GeV, and $|\mu|=400$ GeV are shown in 
Fig.~\ref{muegamma}.
Assuming no accidental cancellation among different diagrams, the 
$\mu \to e\gamma$ constraint requires $\frac{\Delta
m_{12}^2}{\bar{m}_{12}^2}\sin 2\theta_{12} \lesssim 10^{-2}$
for small $\tan\beta$, and about one order of magnitude stronger
for large $\tan\beta$ ($\sim 50$). Therefore, if the mixing angle
$\theta_{12}$ is not very small, $\tilde{e}$ and $\tilde{\mu}$
have to be quite degenerate in order to suppress the contribution to
$\mu \to e\gamma$ by the superGIM mechanism.

Similarly, nonzero $W_{32}$ and $W_{31}$ can contribute to the
rare decays $\tau \to \mu \gamma$ and $\tau \to e \gamma$. The 
current bounds $B(\tau \to \mu \gamma) < 2.9\times 10^{-6}$,
$B(\tau \to e \gamma) < 2.7\times 10^{-6}$~\cite{CLEO}, however,
do not constrain the corresponding mixing angles and mass splittings
for small $\tan\beta$, and constrain them only weakly for large
$\tan\beta$. While $W_{32}$ and $W_{31}$ are not constrained individually,
their product is constrained by $\mu \to e \gamma$ if large splitting
between $\tilde{\tau}$ and $\tilde{\mu}$, $\tilde{e}$ masses exist,
because $\mu \to e\gamma$ can occur through the $\tilde{\tau}$ loop.

At the muon collider, an extremely intense very low energy muon
source will be available for greatly improving the search for
rare muon decays. It is estimated that the bound on 
$B(\mu \to e \gamma)$ may be pushed down to $\sim 10^{-14}$ and even
better for $\mu \to e$ conversion~\cite{Marciano}. This will
dramatically improve the probing range of
$\frac{\Delta m_{12}^2}{\bar{m}_{12}^2}$ and $\sin 2\theta_{12}$.
The discovery of $\mu \to e \gamma$ will have important implications
on the flavor structure and hence will be extremely exciting.
However, the prediction of $B(\mu \to e\gamma)$ in SUSY theories
still depends on many parameters and there are many diagrams which
may add up or cancel each other. A single number $B(\mu \to e\gamma)$
is not enough for us to understand the whole flavor mixing matrices.
We would like to get more handles on the $W$ matrices. At the
muon collider, in addition to measuring $B(\mu \to e\gamma)$,
if sleptons can be produced, we can probe these flavor mixing
matrices directly from the flavor-changing slepton production and decay.
They have simpler dependence on the SUSY parameters and there
can be different modes to be measured. Therefore, they may provide
more and clearer information for the flavor mixing matrices

\begin{figure}
\centerline{\psfig{file=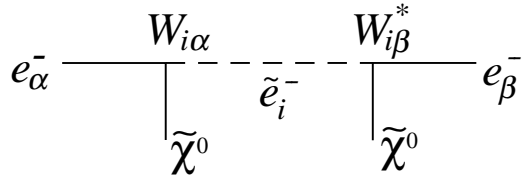,width=0.6\textwidth}}
\caption{The flavor-violating process involving a single
  slepton production.}
\label{lfvgraph}
\end{figure}

We now consider the flavor-violating signals from on-shell slepton
production at the muon collider. The signals we look for consist of
a pair of unlike flavor leptons in the final state, $\mu^+ \mu^-
\to e_{\alpha}^+ e_{\beta}^- \tilde{\chi}^0 \tilde{\chi}^0$.
For simplicity, let us consider the flavor-violating processes
involving a single slepton\footnote{The formalism for correlated slepton
pair production is more complicated and is presented in~\cite{ACFH2}.
However, the essential results remain unchanged.}
as shown in Fig.~\ref{lfvgraph}.
Summing over the amplitudes of different flavor sleptons, the 
cross section is proportional to
\begin{equation}
\sum_{ij} W_{i\alpha} W^*_{i\beta} W_{j\alpha}^* W_{j\beta} 
\frac{1}{1+ix_{ij}} \, , 
\end{equation}
where $x_{ij}\equiv \Delta
m_{ij}/\Gamma$, and $\Gamma$ is the slepton decay width.
For two generation mixing, it reduces to (1-2 mixing only)
\begin{equation}
 \sin^2 2\theta_{12} \frac{x_{12}^2}{2(1+x_{12}^2)} \, .
\end{equation}
As with the low-energy signal, the flavor -violating collider
signal vanishes in the limit of degenerate sleptons. However,
the collider signal is suppressed only for $\Delta m < \Gamma$
where the quantum interference between different flavor sleptons
becomes important, in constrast with the low energy signal which
is suppressed by $\Delta m/\bar{m}$.

\begin{figure}
\centerline{\psfig{file=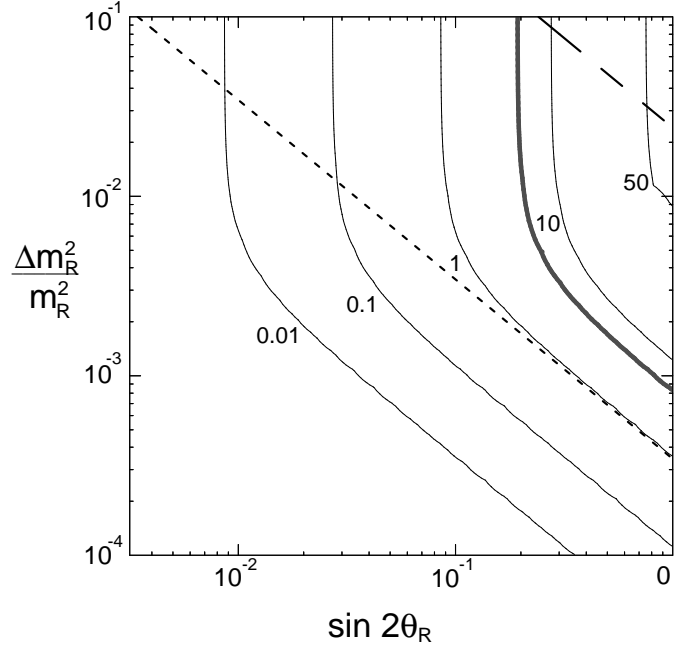,width=0.7\textwidth}}
\caption{Contours of constant 
$\sigma (\mu^+\mu^-\to e^{\pm} \mu^{\mp} \tilde{\chi}^0\tilde{\chi}^0 )$
(solid) in fb for $\protect\sqrt{s} = 500 \text{ GeV}$,
$m_{\tilde{e}_R}, m_{\tilde{\mu}_R} \approx 200 \text{ GeV}$, and $M_1
= 130 \text{ GeV}$.  The thick contour represents the
experimental reach with integrated luminosity 20 fb$^{-1}$.
Constant contours of $B(\mu \to e\gamma)=4.9\times 10^{-11}$ (dashed)
and $10^{-14}$ (dotted) are also plotted for
$m_{\tilde{l}_L} \approx 350$ GeV.}
\label{lfvplot}
\end{figure}

Having discussed the flavor-violating cross sections, we now
consider the reach in the flavor mixing parameter space at the
muon collider. We consider a case with the following SUSY parameters:
$m_{\tilde{e}_R}, m_{\tilde{\mu}_R} \approx 200$ GeV, $M_1=M_2/2
=130$ GeV, $\mu=-400$ GeV, and $\tan\beta=2$. The LSP is almost
pure bino and the cross section has little dependence on $\mu$,
$M_2$, and $\tan\beta$ in this region. For center of mass energy
$\sqrt{s}=500$GeV, we calculate the cross section of the flavor-violating
signal $\mu^{\pm} e^{\mp} \tilde{\chi}^0 \tilde{\chi}^0$ as a
function of $\sin 2\theta_{12}$ and $\Delta m^2$, and the result is 
shown in Fig.~\ref{lfvplot}.
The major backgrounds come from $WW$, $W\nu \mu$, and $\tau\tau$
events. Assuming the similar cuts in the Next Linear Collider 
studies can be applied,
the backgrounds can be reduced to $\sim 5$ fb while keeping about
30\% of the signals. With the integrated luminosity 20 fb$^{-1}$,
the 3$\sigma$ discovery limit is then $\sigma \sim 5$ fb, which
is shown by the thick contour in Fig.~\ref{lfvplot}. We can see
that $\sin \theta_{12}$ may be probed to $\sim 0.1$ for $\Delta m
> \Gamma$. For comparison, we also superimpose the contours of the
current $B(\mu \to e\gamma)$ constraint and the expected reach
by the intense muon source at the muon collider front end. The collider
probe reach far below the current constraint for small $\Delta m$,
but not as far as the expected $\mu \to e\gamma$ reach. However,
$\mu \to e\gamma$ could receive contributions from many diagrams and
has complicated dependence on many SUSY parameters. Therefore, it
is more difficult to be disentangled to give precise information
of flavor mixings than the collider signal.

It is also possible to probe 23 and 13 mixings by looking at
final states of various flavor leptons. For the 23 mixing,
the analysis is similar except that we look for final states with
a $\tau$ instead of an $e$. Only hadronic decays of $\tau$ can
be used as signals so the reach of the 23 mixing is a little
worse than the 12 mixing. Nevertheless, it is still very interesting
compared with no constraint from the current $\tau \to
\mu \gamma$ bound. For the 13 mixing, because the dominant
$t$-channel contribution to the slepton production 
always involves the initial state
muons, we can only probed it (if possible) through the smaller
$s$-channel contribution. Therefore, the probing power is not
promising. In constrast, the electron collider can probe 13
mixing quite well, but not the 23 mixing.

\begin{figure}
\centerline{\psfig{file=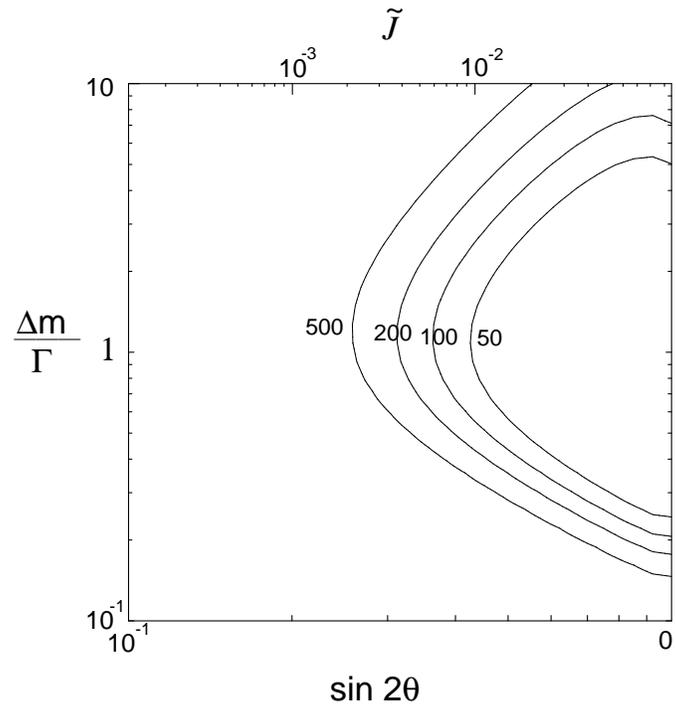,width=0.7\textwidth}}
\caption{3$\sigma$ slepton CP violation
discovery contours for the integrated luminosity given (in fb$^{-1}$).
The CP-violating phase is fixed to $\sin\delta = 1$.  The SUSY
parameters are as given in the text.}
\label{cpvplot}
\end{figure}

The flavor mixing matrices may also contain CP-violating phases.
In the presence of CP violation, the cross sections 
$\sigma_{e_{\alpha}^+ e_{\beta}^-}$ and $\sigma_{e_{\beta}^+ e_{\alpha}^-}$
are no longer equal~\cite{ACFH2,Keung}, and the difference is proportional
to the SUSY analogue to the Jarlskog invariant, $\tilde{J}$, which is
defined by
\begin{equation}
Im \left[ W_{i\alpha} W^*_{i\beta} W_{j\alpha}^* W_{j\beta} \right] 
= \widetilde{J} \sum_{k\gamma} \varepsilon_{ijk}
\varepsilon_{\alpha\beta\gamma} \, . 
\label{J}
\end{equation}
Therefore, the asymmetry between the numbers of $\mu^+ e^-$
and $e^+ \mu^-$ events provides a signal for the CP violation
in the slepton flavor mixing matrices. This CP-violating
asymmetry can only be significant for large 3 generation
mixings and the mass splittings among different generations
comparable to $\Gamma$. Therefore, if it is discovered, it
points toward a very specific flavor structure. Fig.~\ref{cpvplot}
shows the 3$\sigma$ discovery limit of the CP asymmetry for
the set of SUSY parameters: $m_{\tilde{l}_R}=150$ GeV,
$M_1=M_2/2=100$ GeV, $\mu=-400$ GeV, $\tan\beta=2$, and assuming
$\Delta m_{12}=\Delta m_{23}\equiv \Delta m$, $\theta_{12}=
\theta_{23}=\theta_{13}\equiv \theta$, $\sin \delta =1$ in the
standard parametrization~\cite{Chau}.

In conclusion, if supersymmetry is discoverd, there will be
a long and exciting road to measure the SUSY parameters
in order to understand the underlain theory in nature.
The SUSY flavor mixing matrices may provide important clues
to the flavor structure and hence should be an important
subject to study. At the muon collider, the flavor violation
can be probed both by the low energy processes at the front
end and direct slepton production at the collider. 
They can provide complementary information on the flavor mixings 
of the underlain SUSY theory.




\begin{thebibliography}{99}

\bibitem{susytest}
J. L. Feng, H. Murayama, M. E. Peskin and X. Tata, {\it Phys. Rev.}
{\bf D52}, 1418 (1995);
M. M. Nojiri, K. Fujii and T. Tsukamoto, {\it Phys. Rev.}
{\bf D54}, 6756 (1996);
H.-C. Cheng, J. L. Feng and N. Polonsky, {\it Phys. Rev.}
{\bf D56}, 6875 (1997), and Fermilab-PUB-97/205-T, hep-ph/9706476,
to be published in {\it Phys. Rev.} {\bf D}.


\bibitem{sumrules} 
S. Martin and P. Ramond, {\it Phys. Rev.} {\bf D48}, 5365 (1993);
Y. Kawamura, H. Murayama and M. Yamaguchi, {\it Phys. Lett.}
{\bf B324}, 52 (1994);
H.-C. Cheng and L. J. Hall, {\it Phys. Rev.} {\bf D51}, 5289 (1995);
S. Dimopoulos, S. Thomas and J. D. Wells, {\it Nucl. Phys.}
{\bf B488}, 39 (1997).

\bibitem{ACFH1} N. Arkani-Hamed, H.-C. Cheng, J.L. Feng and L.J. Hall,
  {\it Phys. Rev. Lett.} {\bf 77}, 1937 (1996).

\bibitem{ACFH2} N. Arkani-Hamed, H.-C. Cheng, J.L. Feng and L.J. Hall,
  {\it Nucl. Phys.} {\bf B505}, 3 (1997).

\bibitem{Bolton} R.D. Bolton, {\it et al.}, 
  {\it Phys. Rev.} {\bf D38}, 2077 (1988).

\bibitem{Marciano}  W. Marciano, in these proceedings.

\bibitem{CLEO}
  CLEO Collaboration, K.~W.~Edwards {\em et al.}, {\it Phys.~Rev.} {\bf D55},
  3919 (1997).

\bibitem{Becker} R. Becker and C. Vander Velde, in {\em Proceedings of
  the European Meeting of the Working Groups on Physics and Experiments
  at Linear $e^+ e^-$ Colliders,} ed. P.M. Zerwas, Report No.
  DESY-93-123C, p. 457.

\bibitem{Keung} D. Bowser-Chao and W.-Y. Keung, {\it Phys. Rev.}
  {\bf D56}, 3924 (1997).

\bibitem{Chau}
L.-L.~Chau and W.-Y.~Keung, {\it Phys. Rev. Lett.} {\bf 53}, 1802 (1984).

\end{thebibliography}
\end{document}